%% file: main.tex
\documentclass[letterpaper, 10 pt, conference, review]{ieeeconf}  

\IEEEoverridecommandlockouts                              
\usepackage{graphicx} 
\usepackage{amsmath} 
\usepackage{amssymb}  
\usepackage{xcolor}

\usepackage{algorithm}
\usepackage{algpseudocode}

\usepackage{tikz}
\usetikzlibrary{automata, positioning, arrows}

\usepackage{lineno}

\newtheorem{definition}{Definition}
\newtheorem{theorem}{Theorem}

\DeclareMathOperator{\LimAvg}{LimAvg}

\newcommand{\Xb}{\mathbf{X}} 
\newcommand{\Yb}{\mathbf{Y}}

\newcommand{\cX}{\mathcal{X}} 
\newcommand{\cS}{\mathcal{S}} 
\newcommand{\cY}{\mathcal{Y}} 
\newcommand{\cE}{\mathcal{E}} 
\newcommand{\cH}{\mathcal{H}} 
\newcommand{\oepsi}{\overline{\epsilon}}

\title{\LARGE \bf
Data-driven Abstractions with Probabilistic Guarantees for Linear PETC Systems
}

\author{Andrea Peruffo and Manuel Mazo Jr. 
\thanks{This work was supported by the European Research Council through the SENTIENT project (ERC-2017-STG \#755953).}
\thanks{A. Peruffo and M. Mazo are with Faculty of Mechanical, Maritime and Materials Engineering, 
        TU Delft, Delft, The Netherlands.
        {\tt\small a.peruffo@tudelft.nl}}%
}

\begin{document}

\maketitle
\thispagestyle{plain}
\pagestyle{plain}

\begin{abstract}
    We employ the scenario approach to compute probably approximately correct (PAC) bounds on the average inter-sample time (AIST) generated by an unknown PETC system, based on a finite number of samples. 
    We extend the scenario approach to multiclass SVM algorithms in order to construct a PAC map between the concrete, unknown state-space and the inter-sample times.
    We then build a traffic model applying an $\ell$-complete relation and find, in the underlying graph, the cycles of minimum and maximum average weight: these provide lower and upper bounds on the AIST.
    Numerical benchmarks show the practical applicability of our method, which is compared against model-based state-of-the-art tools.
\end{abstract}


\section{INTRODUCTION}
\label{sec:intro}

In the last decades, thanks to the increasing digitalisation and use of communication networks, control systems have faced several new challenges. Arguably, the most compelling task is ensuring the stability of an interconnection between an analog plant and a digital controller, possibly in spite of imperfect communication means.
The interface between a (digital) controller and an (analog) plant is typically implemented with a periodic sampling of the plant, whose measurements are transmitted to the controller, which computes the actions in order to optimise a performance cost. 
%
The sampling period itself represents a tradeoff between performance and energy consumption -- any control signal entails an action by the actuator, besides potentially congesting the communication network. 
%

Event-triggered control (ETC) is a paradigm that tackles this issue, adjusting the sampling according to the satisfaction of a condition, whilst maintaining the stability guarantees. 
This notion has originally been developed in \cite{astrom2002comparison}, and further notably developed in \cite{tabuada2007event} where guarantees on the closed loop performance are ensured.
A practical implementation procedure of this scheme is the periodic ETC (PETC), where a stability condition is periodically checked whilst the measurement is transmitted solely when the condition is verified. 
Whilst the advantage of this approach is apparent, formally and quantitatively \emph{measuring} its performance has been tackled only recently.

The use of formal abstractions for PETC models has been preparatory in this sense, see e.g. \cite{kolarijani2016formal, tabuada2009verification}.
More recently, 
the work \cite{gleizer2020scalable} provides a characterisation in terms of traffic abstractions: a finite-state automaton considering the inter-sample times (ISTs) sequences of a PETC system.
The procedure is applied in \cite{gleizer2020towards}, shifting the focus to $\ell$ consecutive inter-sample times: this offers a conservative estimate of the long-term performance of a PETC system.
Further, in 
\cite{de2021computing} the authors build a traffic abstraction in order to compute the minimum average inter-sample time (AIST) of a PETC system, a metric that can be translated directly to the expected network load or resource utilization. 
ETCetera, a state-of-the-art tool to compute the sampling performance of ETC systems is presented in \cite{etcetera}.

We build upon this literature, aiming at constructing finite-state abstraction of a PETC system, with a crucial difference.
In this work, we do not require any knowledge about the underlying concrete system.
Nevertheless, we construct a data-driven map between the concrete state-space and the inter-sample times with 
probably approximately correct (PAC) guarantees, which acts as the foundation for a traffic
 abstraction to compute reliable bounds on the AIST of the unknown system.
Data-driven abstractions have recently gained interest: in \cite{cubuktepe2020scenario, badings2021sampling} the authors provide a sampled-based interval MDP, employing the scenario approach to provide bounds on the transition probabilities, whilst the state space is partitioned with a gridding procedure.

Since the concrete model is unknown, building the relationship between the state-space and the corresponding ISTs is a non-trivial challenge. We provide a map based on a finite number of samples (often called scenarios) which we assume can be computed via simulations. 
We adapt the scenario approach 
which provides probabilistic guarantees by solving a convex optimisation program \cite{campi2008exact, campi2018general, garatti2019risk, campi2020scenario}. 
We use this map to 
partition the state space of the concrete system. Each partition corresponds to an abstract state of the traffic model, whose transitions are governed by the so-called domino rule. 
We then extrapolate upper and lower bounds on the AIST; by increasing $\ell$, the procedure can be refined until the upper and lower bounds reach a desired precision.
The bounds are PAC, as the true AIST value is contained within these boundaries  with a probability not less than a user-designed confidence level.

\textbf{Contributions}
In this work, we elaborate an extension to \cite{campi2020scenario} 
as we tailor the scenario approach to multiclass SVM algorithms. 
%
We employ this result to build a data-driven abstraction of an unknown PETC system with probabilistic guarantees of correctness. 
%
We aim at estimating their  resource consumption by identifying the possible (infinite) sequences of inter-sample times. To this end, we provide probably correct bounds that tightly approximate the desired consumption cost. Finally, we compare our procedure against the model-based tool ETCetera \cite{etcetera}.


\section{Finite Abstraction of PETC via TS}
\label{sec:prob-formulation}

Consider a linear time-invariant plant controlled with sample-and-hold state feedback described by
\begin{equation}
\label{eq:or-linear-sys}
    \dot{\xi} (t) = A \xi (t) + B \nu (t), 
    \quad
    \nu (t) = K \hat{\xi} (t),
\end{equation}
where $\xi(t) \in \mathbb{R}^{n_x}$ is the plant’s state with initial value $x_0 := \xi(0)$, $\hat{\xi}(t) \in \mathbb{R}^{n_x}$ is the state measurement available to the controller, $\nu(t) \in \mathbb{R}^{n_u}$ is the control input, $n_x$ and $n_u$ are the state-space and input-space dimensions, respectively, and $A$, $B$, $K$ are matrices of appropriate dimensions. The measurements are updated to the controller only at specific sampling times, with their values being zero-order held on the controller: let $t_i \in \mathbb{R}_+$, $i \in \mathbb{N}_0$ be a sequence of sampling times, with $t_0 = 0$ and $t_{i+1} - t_i > \varepsilon$ for some $\varepsilon > 0$; then $\xi(t) = \xi(t_i)$, $\forall \, t \in [t_i,t_{i+1})$.
In ETC, a triggering condition determines the sequence of times $t_i$. In the case of PETC, this condition is checked only periodically, with a fundamental checking period $h$.

We typically consider the family of quadratic triggering conditions \cite{heemels2012periodic} where we additionally set a maximum inter-sampling time $\overline{\kappa}$, known as the \textit{heartbeat} of the system. Formally, we define the $(i+1)$-th triggering time as
\begin{equation}
\label{eq:trigger-cond}
    t_{i+1} = 
    \\
    \inf
    \left\{
        k h > t_i \ \left| \ 
        \begin{aligned}
        &
        \begin{bmatrix}
        \xi(k h) \\ \xi(t_i)
        \end{bmatrix}^T
        R 
        \begin{bmatrix}
        \xi(k h) \\ \xi(t_i)
        \end{bmatrix} 
        > 0 
        \\
        &
        \text{ or } k h - t_i \geq \overline{\kappa}
        \end{aligned}
        \right.
    \right\},
\end{equation}
where  $k \in \mathbb{N}$ and $R \in \mathbb{S}^{2n_x}$ is the designed triggering matrix, and $\overline{\kappa}$ is the chosen maximum inter-sample time. 
Every run of system \eqref{eq:or-linear-sys}, starting from initial condition $\xi(0)$, generates an infinite sequence of samples $\{\xi(t_i)\}$ and of inter-sample times $\{ \tau_i \}$. 
In the following, we solely consider PETC settings hence $\tau \in  h \cdot \{ 1, 2, \ldots \overline{\kappa}\}$; 
for simplicity and without loss of generality, we consider $h=1$.

A typical metric of interest is the \textit{average inter-sample time} (AIST), defined for every initial state $\xi(0)$ as 
\begin{equation}
\label{eq:aist-definition}
    \text{AIST}(\xi(0)) = 
    \liminf\limits_{n\rightarrow \infty} \frac{1}{n+1} \sum_{i=0}^n \tau(\xi(t_i)).
\end{equation}
Notice that we use the liminf instead of lim in case the regular limit does not exist \cite{de2022chaos}, making the AIST a well-defined metric.
Additionally we define the \textit{smallest} and the \textit{largest} average inter-sample time (SAIST and LAIST, respectively) as 
\begin{equation}
\label{eq:saist-definition}
    \text{SAIST} := \inf_{\xi \in \mathbb{R}^{n_x}} \liminf\limits_{n\rightarrow \infty} \frac{1}{n+1} \sum_{i=0}^n \tau(\xi(t_i)),
\end{equation}
\begin{equation}
\label{eq:laist-definition}
    \text{LAIST} := \sup_{\xi \in \mathbb{R}^{n_x}} \limsup\limits_{n\rightarrow \infty} \frac{1}{n+1} \sum_{i=0}^n \tau(\xi(t_i)).
\end{equation}
Both the SAIST and the LAIST are crucial metrics to assess the performance (in terms of resources consumption or band utilisation) of an ETC implementation. However, the present expressions \eqref{eq:saist-definition}, \eqref{eq:laist-definition} require the search for a state $\xi$ over $\mathbb{R}^{n_x}$ and the choice of a sufficiently large $n$, which are tasks computationally hard (if at all possible). 
To overcome this impediment, we approach this problem exploiting the (finite-state) abstractions theory. 

\subsection{Abstractions and Transition Systems}
\label{subsec:tr-sys-and-abs}

The abstraction of a system allows the analysis of large (even infinite) models.
We usually consider a \textit{concrete} model, which is treated as a ground truth; this is transformed, simplified, and adapted to accommodate the analysis of its behavior. 
Several kinds of abstractions are available for different purposes: we employ a finite-state abstraction in the form of a (weighted) transition system (WTS).
\begin{definition}[Weighted Transition System (adapted from  \cite{tabuada2009verification,chatterjee2010quantitative})]
    A transition system $\cS$ is a tuple $(\cX,\cX_0,\cE,\cY,\cH, \gamma)$ where:
\begin{itemize}
    \item $\cX$ is the set of states,
    \item $\cX_0 \subseteq \cX$ is the set of initial states,
    \item $\cE \subseteq \cX \times \cX$ is the set of edges, or transitions, 
    \item $\cY$ is the set of outputs, and
    \item  $\cH : \cX \rightarrow \cY$ is the output map, 
    \item $\gamma$ : $\cE \rightarrow \mathbb{Q}$ is the weight function.
\end{itemize}
\end{definition}
The original definition also includes the action set 
 $\mathcal{U}$ 
which is here omitted since we solely consider autonomous systems; we further consider finite-state systems, where the cardinality of $\cX$ is finite. We tacitly consider \textit{non-blocking} transition systems, i.e. automata where every state is equipped with at least one outgoing transition.

Let us define $r = x_0 x_1 x_2 \ldots$ an infinite internal behavior, or \textit{run} of $\cS$ if $x_0 \in \cX_0$ and $(x_i,x_{i+1}) \in \cE$ for all $i \in \mathbb{N}$, and, with slight abuse of notation, $\cH(r) = y_0 y_1 y_2 \ldots$ its corresponding \textit{external} behavior, or trace, if $\cH(x_i) = y_i$ for all $i \in \mathbb{N}$. Similarly, we define $\gamma(r) = v_0 v_1 \ldots$ the sequence of weights from run $r$, where $\gamma(x_i, x_{i+1}) = v_i$.
%
Let us further consider a value function, i.e. a function mapping  an infinite sequence of weights to a (possibly finite) value, as 
\begin{equation}
\label{eq:lim-avg-def}
    \LimAvg(\gamma(r)) := \liminf\limits_{n \rightarrow \infty} \frac{1}{n+1} \sum_{i=0}^n v_i .
\end{equation}
A weighted automaton equipped with the LimAvg value function is called a LimAvg-automaton \cite{chatterjee2010quantitative}. 

%
Closely related to this metric, we can define 
\begin{eqnarray}
\label{eq:sup-lim-avg}
    \overline{V}(\cS) := \sup\{\LimAvg(\gamma(r)) \mid r \text{ is a run of } \cS\},
    \\
    \underline{V}(\cS) := \inf\{\LimAvg(\gamma(r)) \mid r \text{ is a run of } \cS\}.
\end{eqnarray}
Remarkably, \cite{chatterjee2010quantitative,karp1978characterization} show that we can recover from a WTS the value of $\underline{V}(\cS)$ and $\overline{V}(\cS)$ in polynomial time.
The value function in \eqref{eq:sup-lim-avg} evaluates to the \textit{smallest average cycle} (SAC) of the graph underlying the transition system. 
In many contexts, the performance of systems as the (minimum) average resource consumption is modeled as the minimum cycle mean problem.
Analogously, we may be interested in the \textit{largest} average cycle (LAC) to analyse the maximum average resource consumption. 
%
%


We ideally would abstract a PETC system as a WTS. However, the abstraction requires the knowledge on the internal behaviour of the system, i.e. the states $\xi$ and their dynamics.
%
We then introduce the notion of \emph{behavioural inclusion}: the abstraction is oblivious to the internal behaviour of a system, but we require that all traces observed in the concrete system are also observed in the abstraction. 
In practical terms, 
the ISTs of any trajectory of the concrete PETC system must overlap the output of a run of the weighted TS.


%
A natural way of building a {behavioural} {inclusion} abstraction is by mapping each possible output symbol $\tau_i$ to an abstract state $x_i$. We may elaborate this intuition through a so-called $\ell$-complete model:
\begin{definition}[(Strongest) $\ell$-complete abstraction \cite{de2021computing}]
\label{def:sl-ca}
Let $\cS :=(\cX,\cX_0,\cE,\cY,\cH)$ be a transition system, and let $\cX_\ell \subseteq \cY_\ell$ be the set of all $\ell$-long subsequences of all behaviors in $\cS$. Then, the system $\cS_\ell = (\cX_\ell, B_\ell(\cS), \cE_\ell, \cY, \cH)$ is called the (strongest) $\ell$-complete abstraction (S$\ell$-CA) of $\cS$, where
\begin{itemize}
    \item $\cE_l$ = $\{(k \sigma, \sigma k') \, | \, k,k' \in \cY$, $\sigma \in \cY_{\ell-1}$,  $k \sigma$, $\sigma k'$ $\in \cX_\ell$ \},
    \item $\cH(k \sigma) = k$,
\end{itemize}
\end{definition}
where we denote $B_\ell(\cS)$ all the possible (infinite) external traces of system $\cS$.
The intuition behind the S$\ell$-CA is to encode each state as an $\ell$-long external trace. Notice that the transitions follow the so-called ``domino rule'': e.g., let us assume $\ell=3$ and let us observe a trace \textit{abc} (this trace corresponds to a single abstract state);
the next $\ell$-trace must begin with \textit{bc}. 
Therefore, a transition from \textit{abc} can lead to 
e.g. \textit{bca}, \textit{bcb}, \textit{bcc}.
Finally, the output of a state is its first element: state \textit{abc} has output $\cH(abc) = a$.

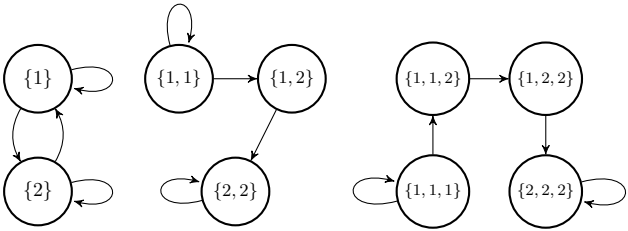
\begin{figure}
    \centering
    \input{models/two_ist_ts}
    \caption{S$\ell$-CA with $\tau_i=\{1,2\}$ and $\ell$=1 (left), $\ell=2$ (center), $\ell=3$ (right).}
    \label{fig:slca-example}
\end{figure}


The S$\ell$-CA translation is particularly useful when little or no information is available about the concrete system, as only the external behaviour, i.e. the sequences of $\tau_i$, is employed to build the abstraction. 
%
In order to compute the AIST, which is a function of $\xi(0)$, we shall need a map, denoted $\mathcal{T}: \xi \rightarrow \tau$, that relates concrete states to inter-sample times in order to compute the initial abstract state $x_0$.
Further, the S$\ell$-CA is in general non-deterministic: 
we cannot evaluate exactly the LimAvg (as every run starting from $x_0$  branches into several possible runs), thus we compute the SAC and LAC (see \eqref{eq:sup-lim-avg}) and use them as upper and lower approximations of the LimAvg, i.e. the AIST. 
Figure \ref{fig:slca-example} shows three examples of S$\ell$-CA, where $\tau_i = \{1,2\}$ and $\ell=1,2,3$. Considering $\ell=3$ and as initial state $x_0 = \{1,1,1\}$, we notice that its SAC and LAC are the self-loop on states $(1,1,1)$ and $(2,2,2)$ respectively. Therefore, the AIST is within $[1,2]$.
%
%

The triggering condition (see \eqref{eq:trigger-cond}) allows solely a \emph{subset} of all possible $\ell$-sequences of ISTs.
%
A model-based method can exploit the knowledge of the concrete system to check which sequences of length $\ell$ are admissible and construct a tailored TS.

In the following, we employ a scenario approach to obtain the map $\mathcal{T}$ with guarantees of correctness and carefully tailor the abstract state space without any knowledge of the underlying concrete system.


\section{Data-driven Abstractions with Guarantees}
\label{sec:scenario-approach}

Naturally, full knowledge of the concrete system is useful to compute the mapping $\mathcal{T}$  between the states $\{ \xi(t_i) \}$ and the output traces $\{ \tau_i \}$. 
However, in many applications this knowledge may be unavailable, unreliable, or simply expensive to obtain. 
To overcome this impediment, we employ the scenario approach \cite{campi2008exact}.


Let us assume we collect $N$ samples $Q_i = (\Xb_i, \Yb_i)$, $i = 1, \dots N$, where $\Xb_i$ represents the $i$-th sample of the concrete system state and $\Yb_i$ represents an $\ell$-sequence of ISTs.
Since $\Yb_i$ have finite cardinality, we may associate them with \textit{classes} or \emph{labels} from a machine learning viewpoint. 
From now on, we consider $\{ \Yb_i \}$ belonging to $L$ different classes, i.e. $\Yb_i \in \{1, \dots L\}$.

\begin{figure}
    \centering
    \includegraphics[width=1.05\linewidth]{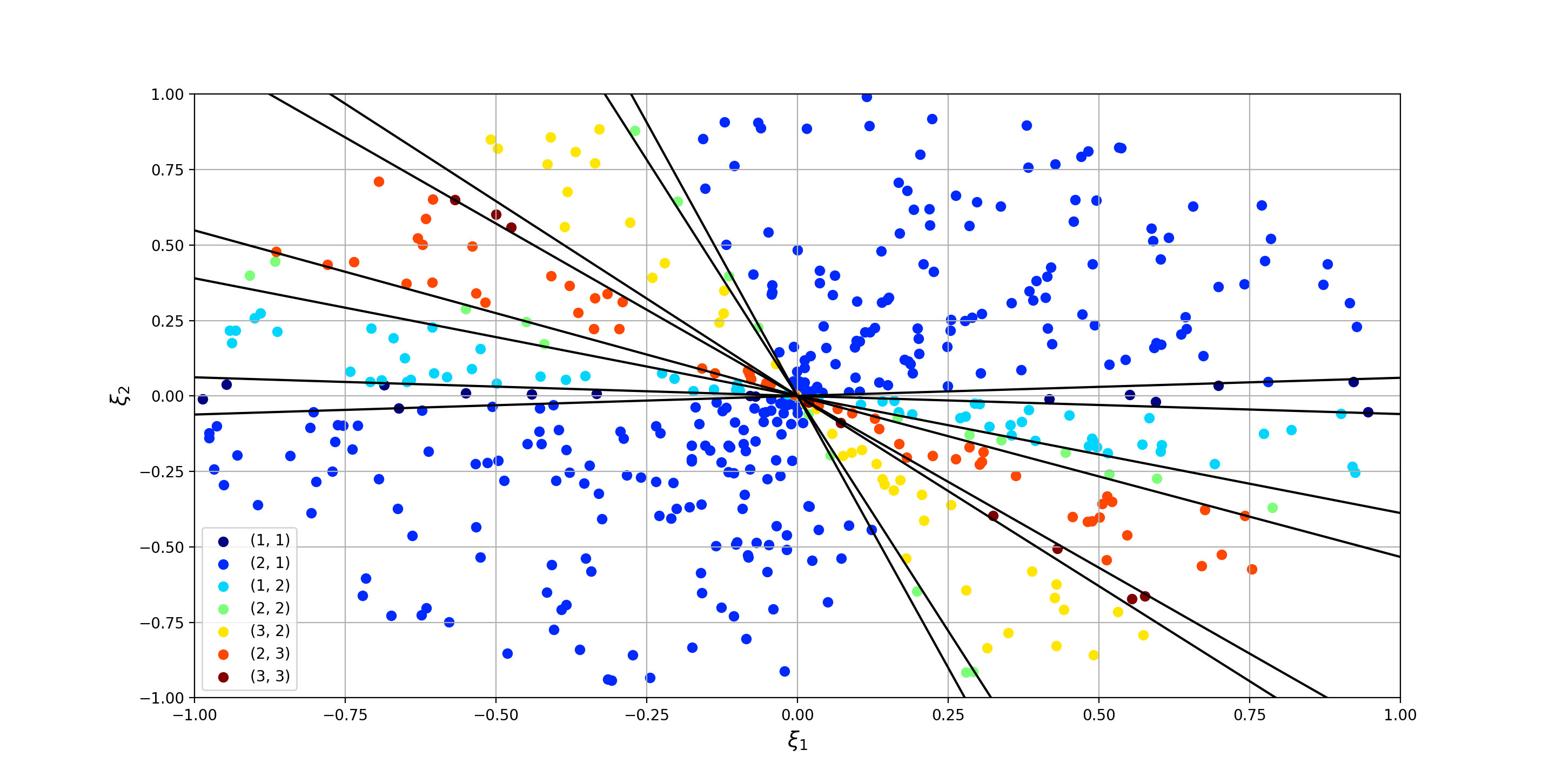}
    \caption{IST regions are unions of (diametrically opposite) cones: black lines denote the regions' boundaries, depicted along with 1000 samples in different colors, for $\ell=2$.}
    \label{fig:l-sequences}
\end{figure}

For PETC systems, the regions of the state space that correspond to $\ell$-sequences of ISTs have a well-defined shape, as they are unions of cones \cite{gleizer2020scalable, gleizer2020towards, de2021computing}, as depicted in Fig. \ref{fig:l-sequences}. 
We shall then use this intuition to partition the state space and construct the abstract states.
%
This allows to reduce the number of abstracted states in comparison to an agnostic gridding procedure. 
%

\subsection{Linear Separability with Veronese Embedding}
\label{subsec:veronese}

We draw inspiration from the SVM literature, where hyper-planes are constructed in order to split linearly separable datasets. 
Typically, when the samples are not linearly separable, we employ the kernel trick \cite{weston1998multi} to embed the data into a higher-dimensional space, where a linear separation is possible. 
A model given by \eqref{eq:or-linear-sys}-\eqref{eq:trigger-cond} separates the state space into (union of) cones: hence, a conic transformation to the samples $\mathbf{X}_i$ grants a linearly separable dataset. 
This embedding is known as the Veronese map \cite{mazo2009self} of order 2 , $\mathcal{V}_2: \mathbb{R}^n \rightarrow \mathbb{R}^{\frac{n(n+1)}{2}}$, defined by 
\begin{equation}
    \mathcal{V}_2(x) := [x_1^2 \ x_1 x_2 \ x_2^2 \ \ldots \ x_n^2],
\end{equation}
where $x_j$ indicates the $j$-th component of vector $x$.
This embedding represents a tailored kernel trick for the problem at hand.
Further, 
the conic partitions fulfil an \emph{encapsulation} constraint: 
every point of the concrete system $\xi$ whose next IST is $j$ satisfies
\begin{equation}
\label{eq:encapsulation-cones}
    \xi^T N_{i} \xi < 0 \text{ for }i < j, \quad \xi^T N_j \xi > 0,
\end{equation}
where the set of matrices $N_i$ is a function of $A$, $B$, $K$ and $R$ (see \eqref{eq:or-linear-sys}-\eqref{eq:trigger-cond}).
Intuitively, the $N_i$ indicate a violation of the $i$-th triggering condition. For instance, if $\xi^T N_1 \xi < 0$, the triggering condition is not violated after 1 inter-sample times; conversely, if $\xi^T N_2 \xi > 0$ the triggering condition is violated after 2 ISTs, provided that it is not violated after 1 IST (i.e. $\xi^T N_1 \xi < 0$ must hold).

We map conditions \eqref{eq:encapsulation-cones} with the Veronese embedding and obtain a set of linear constraints 
\begin{equation}
\label{eq:encapsulation-cones-veronese}
   W_{i} \cdot \mathcal{V}_2(\xi) < 0 \text{ for }i < j, \quad W_j \cdot \mathcal{V}_2(\xi) > 0, 
\end{equation}
where $W_i$ shall be recovered from the samples $Q_i$.
Constraints \eqref{eq:encapsulation-cones-veronese} closely resemble a multiclass SVM approach. In the following, we outline the scenario theory for a (more general) SVM multiclass problem, highlighting the adjustments needed to fit the encapsulation problem \eqref{eq:encapsulation-cones-veronese}.

\subsection{Sample-based Partitioning via Classification}
\label{subsec:multi_SVM}

Building upon \cite{campi2020scenario}, where binary SVM with scenario guarantees is presented, 
we extend this approach to multiclass SVM algorithms.
This procedure provides a classifier with guarantees of correctness, which we denote as the map $\mathcal{T}$.
Two methodologies exist to achieve multiclass classification \cite{hsu2002comparison}: the \emph{one-vs-one} reduction, 
where the multi-class problem is split into multiple binary classification problems, and the \emph{one-vs-all} approach, which defines one hyperplane per class.
Notably, the one-vs-all approach can be written in one single optimisation program: for clarity, we select the 
 Crammer-Singer formulation \cite{crammer2001algorithmic}.
%
The extension to one-vs-one and other one-vs-all formulation proceeds similarly to the following discussion, and is omitted here for brevity.

The one-vs-all SVM algorithm consists of one convex problem considering $L$ (one per class) hyperplanes, defined
\begin{equation}
 f_j(\Xb_i) := W_j \Xb_i + b_j, \text{ for } j \in \{1, \dots, L\},     
\end{equation}
where $W_j$ $\in \mathbb{R}^{n \times 1}$, $b_j \in \mathbb{R}$, $\forall \, j \in \{1, \ldots L\}$.  The classification decision follows the \textit{winner-takes-all} policy; formally
\begin{equation}
\label{eq:multiclass-svm-decision}
    D(\Xb_i) := \arg \max \{ W_j \Xb_i + b_j \}.
\end{equation}
For every sample $\Xb_i$, the corresponding hyperplane is denoted as $(W_{y_i}, b_{y_i})$, i.e. the $\Yb_i$-th hyperplane. 
For a correct classification, the value of hyperplane $W_{y_i} \mathbf{X}_i + b_{y_i}$ ought to be greater than all others: we may rewrite \eqref{eq:multiclass-svm-decision} as 
\begin{equation}
\label{eq:svm-one-hyperplane-greater-others}
\zeta
- (W_{y_i}-W_j) \Xb_i + (b_{y_i}-b_j) \leq 0, 
\quad 
\forall j \neq \Yb_i, \ \forall \, i.
\end{equation}
where the hyper-parameter $\zeta>0$ is added for numerical stability.
%
Conditions \eqref{eq:svm-one-hyperplane-greater-others} can be simplified as 
\begin{equation}
\label{eq:max-constraint-svm}
    g(\Xb_i) := 
    \max_{j \neq y_i} \{\zeta -(W_{y_i}-W_j) \Xb_i - (b_{y_i}-b_j)   \}.
\end{equation}

Constraints  $g(\Xb_i) \leq 0$ can be satisfied only if the dataset is linearly separable:
to further accommodate for more general cases, 
we employ slack variables $\theta_i \geq 0$, $i=1, \ldots, N$, impose $g(\Xb_i) \leq \theta_i$ and minimise the sum of $\theta_i$.
The multi-class SVM problem expression follows standard formulation \cite{weston1998multi}. An $L$-class classifier can be obtained by solving the following convex program:
\begin{equation}
\label{eq:scenario-SVM-problem}
\begin{aligned}
    \min_{W, b, \theta_i \geq 0} \quad 
    &  
    \sum^L_{i=1} ||W_i||^2 + \rho \sum_{i=1}^{N} \theta_i, 
    \\
    \text{s.t.} \quad 
    &
    g(\Xb_i) \leq \theta_i, 
    \quad  \, i = 1, \ldots, N,
\end{aligned}
\end{equation}
where $\rho>0$ is a hyper-parameter balancing the trade-off between the correctness of the algorithm (the number of positive $\theta_i$) and its cost (the norm of matrices $W_j$).
%
%

Let us now introduce a quantitative measure to evaluate the reliability of the classifier. 
We present 
the concept of risk (or violation probability), which is a measure of the probability that a (new, unseen) sample is misclassified \cite{campi2020scenario}. 
\begin{definition}
    The probability (risk) of violating a constraint $f(\Xb_{j}, \Yb_j)$, $j > N$, is denoted
    \begin{equation}
        R(\mathbf{S}) := \mathbb{P}[ (\Xb_{j}, \Yb_j) : 
        g(\Xb_j) > 0 \, ],
    \end{equation}
where $f(\Xb_j)$ is defined according to Eq. \eqref{eq:max-constraint-svm} and 
where we denote as $\mathbf{S}$ the parameters of the SVM algorithm ($W_j$, $b_j$) for all $j = 1, \ldots L$.
\end{definition}

We can provide a quantitative evaluation of the risk within the optimisation context of program \eqref{eq:scenario-SVM-problem}: we mimic the discussion in \cite{campi2020scenario} for binary SVM and adapt it to a multiclass environment.
\begin{theorem}[Adapted from \cite{campi2020scenario}]
\label{theo:quantitative-risk}
Given a confidence parameter $\beta \in [0,1]$ and $N$ samples, 
it holds that
\begin{equation}
    \mathbb{P}^N [
    \underline{\epsilon}(s^*, N, \beta) \leq 
    R(\mathbf{S}^*) \leq \overline{\epsilon}(s^*, N, \beta)
    ]
    > 1 - 3 \beta,
\end{equation}
where $\mathbf{S}^*$ represents the parameters that minimise program \eqref{eq:scenario-SVM-problem} and $s^*$ identifies the number of violated constraints, i.e. the number of samples that return $g(\Xb_i) > 0$. Further, the event of misclassification can be bounded by 
\begin{equation}  
\label{eq:prob-misclassification}
    \mathbb{P}^N [
    \mathbb{P}[
    \text{misclassification}
    ]
     \leq \overline{\epsilon}(s^*, N, \beta)
    ]
    > 1 - 3 \beta.
\end{equation}
The bounds $\underline{\epsilon}$, $\oepsi$ can be found solving a polynomial equation, function of $s^*$, $N$, $\beta$, as outlined in \cite{campi2020scenario}.
\end{theorem}
The proof follows a similar result in \cite{campi2020scenario} and is 
omitted for brevity.
\paragraph*{Remark}
It is worth highlighting that in program \eqref{eq:scenario-SVM-problem}  the violation of a constraint does not imply misclassification: it reports that the evaluation of two hyperplanes differs less than $\zeta$.
As such, misclassification occurs more rarely than constraints violation. 
Therefore, Theorem \ref{theo:quantitative-risk} can solely provide an upper bound on the probability of misclassification.
Notice that $\beta$ is a tunable parameter, which ensures desired performance of the algorithm, since $\overline{\epsilon}$ is a function of $s^*$, $\beta$ and $N$. \hfill $\square$

The hyperplanes defined by the classification algorithm outline the state-space partitions, and each partition corresponds to an abstract state of the $\ell$-complete model.
Given an initial state $\xi(0)$, we employ the classifier to map it to its corresponding abstract state, say $x_0$. 
We then evaluate the SAC and LAC that are reachable from $x_0$ in order to give upper and lower bounds for the AIST($\xi(0)$).

\paragraph*{Remark} Conditions \eqref{eq:encapsulation-cones-veronese} can be fitted in program \eqref{eq:scenario-SVM-problem} by changing the classification decision \eqref{eq:multiclass-svm-decision} to reflect that the \emph{first} violation of the conic constraints defines the class appointment. Formally, 
\begin{equation}
\label{eq:conic-svm-decision}
    D(\Xb_i) := \arg \min \{ W_j \cdot \mathcal{V}_2(\Xb_i) > 0 \}, 
\end{equation}
so that conditions \eqref{eq:svm-one-hyperplane-greater-others} shall be replaced by \eqref{eq:encapsulation-cones-veronese}. We refer to this procedure as conic multiclass SVM (CM-SVM).
\hfill $\square$



\subsection{AIST Evaluation with Guarantees}
\label{subsec:building-abstraction}

The map $\mathcal{T}$, thanks to the scenario theory, is equipped with probabilistic guarantees of correctness, which are delivered to the partitions of the concrete state space and thus to the abstract states of the S$\ell$-CA. 
Its transitions, on the other hand, are correct by design following the domino rule, albeit they might introduce spurious behaviours.

The state space of the $\ell$-complete model raises a question. Assume $\ell=2$, $\tau_i = \{1, 2\}$,
and our dataset presents the 2-sequences $\{(1, 1), (1, 2), (2,2) \}$ but not 
 $(2,1)$. We can gather a larger dataset until we observe the missing sequence; however, this might never happen as this particular 2-sequence  may not be allowed by the triggering condition.
%
%

We synthesise a classifier based upon the dataset at hand: any new sample with an unseen label (not amongst the $L$ labels witnessed within the sample set) is evidently misclassified. 
We can thus rethink the misclassification of Theorem \ref{theo:quantitative-risk} as entailing both a mis-labeling, i.e. the algorithm assigns a wrong label to a sample, and a new labeling, i.e. a sample presents an unseen 
label.
%
In other words, 
once the classification algorithm computes the partitions, three disjoint events may occur to a new sample: 
$a)$ it is correctly labeled; 
$b)$ it is mislabeled (occurring with probability $P_{ML}$);
$c)$ it belongs to an unseen class (with probability $P_{NL}$).
The two latter events entail a misclassification, hence 
 we can write the bounds of Theorem \ref{theo:quantitative-risk} as 
\begin{equation}
\label{eq:three-events-scenario-probability}
\begin{aligned}
&
    \mathbb{P}^N \left[ 
    P_{ML} + P_{NL} 
    \leq \overline{\epsilon}(s^*, N, \beta) 
    \right] > 1-3\beta.    
\end{aligned}
\end{equation}
As such, after collecting $N$ samples, we solely account for the labels that the algorithm witnesses, set $L$ as the number of different \textit{seen} labels, and proceed to solve optimisation problem \eqref{eq:scenario-SVM-problem}. 
Theorem \ref{theo:quantitative-risk} retains its validity also in presence of unseen labels, 
if we consider all samples belonging to any omitted class as misclassified; in practical terms,  we should directly add them to $s^*$. 
Indeed, the scenario approach concerns the number of violated constraints: in principle, we could purposely exclude any label and its corresponding samples and Theorem \ref{theo:quantitative-risk} could be applied as long as we account for these samples as misclassifications.

Once the CM-SVM program \eqref{eq:scenario-SVM-problem} is solved, we use the resulting partitions to assign a label to any (new) sample of interest. The label corresponds to an abstract state, which is used as initial state for the $\ell$-CTA: we then compute the SAC and LAC (see \eqref{eq:sup-lim-avg}) as lower and upper bounds for the corresponding AIST.
The gap between the SAC and LAC offers additional information, as 
 $\delta_{AIST} = LAC - SAC$
defines the precision of our abstraction: the true AIST dwells within the interval $\delta_{AIST}$, therefore the tighter $\delta_{AIST}$ is, the more precise the information we gather.
%
%
If needed, 
we may increase $\ell$ to refine the abstraction.
In this sense, $\ell$ can be seen as a tradeoff: a large $\ell$ provides a larger state space of the S$\ell$-CA, herald of a higher computational cost, along with more precise bounds.
Further, 
in order to provide a global precision of the abstraction,  we define 
 the EAC (expected average cycle) as the average $\delta_{AIST}$ over all states; formally, 
\begin{equation}
    EAC := 
    \frac{1}{|\cX|} \sum_{x \in \cX} \delta_{AIST}(x).
\end{equation}
The EAC metric embodies the degree of uncertainty in the evaluation of the AIST for every possible state of the abstraction.
A smaller EAC ensures a more precise abstraction.
%


\section{Experimental Evaluation}
\label{sec:experiments}




We show the effectiveness of our method considering a 2-D and a 4-D linear system, and comparing against a model-based technique, in order to validate our results and evaluate their precision and computational cost.

We collect $N=10^4$ random samples $Q_i = (\Xb_i, \Yb_i)$, evaluated from the 2-D linear system in \cite{de2021computing,gleizer2020towards}, 
\begin{eqnarray}
\label{eq:experiments-2d-system}
    &
    \dot{\xi}(t) = A \xi(t) + B u(t), 
    \\ \nonumber
    &
    A = 
    \begin{bmatrix}
    0 & 1 \\ -2 & 3
    \end{bmatrix}, 
    \quad 
    B = 
    \begin{bmatrix}
    0 \\ 1
    \end{bmatrix}, 
    \quad 
    K = 
    \begin{bmatrix}
    0 & -5
    \end{bmatrix},
\end{eqnarray}
with triggering condition $|\xi(t) - \hat{\xi}(t)|^2 > \sigma^2 |\xi(t)|^2$, where  $\sigma = 0.1$. 
We obtain a dataset with 3 possible ISTs, i.e. 
$\Yb_i \in \{1, 2, 3\}$.


We fix the hyper-parameters of the optimisation program \eqref{eq:scenario-SVM-problem} as $\beta=10^{-6}$, $\rho = 10^3$.
%
The train data are used to solve the optimisation problem \eqref{eq:scenario-SVM-problem} and obtain the mapping from the concrete states $\xi(\cdot)$ to the abstract states $x$. We collect the number of violated constraints $s^*$ and compute the corresponding $\oepsi$ bound according to \eqref{eq:prob-misclassification}.
%
Table \ref{tab:svm-bounds} reports 
the number of labels $L$, the number of violated constraints $s^*$, its corresponding bound $\oepsi$ and the empirical percentage of violated constraints (computed on additional $10^4$ samples) and finally the computational time of the CM-SVM algorithm, 
 for different combinations of $\ell$ and $\sigma$.
%
As expected, the empirical constraint violation percentage is smaller than $\overline{\epsilon}$ for all experiments.
%

\begin{table}[]
    \centering
    \caption{Performance of the CM-SVM algorithm.
    }
    \label{tab:svm-bounds}
    \begin{tabular}{c|ccccccc}
          & 
          &  &  &  & Empir. &  CM-SVM
         \\
         $\sigma$ & 
         $\ell$ & $L$ & $s^*$ & $\overline{\epsilon}$ & Violat. & time [s] 
         \\ \hline
         0.1 
         & 1 & 3 & 99 & 0.020 & 1.14\% & 1.4 
         \\
         & 5 & 19 & 212 & 0.034 & 2.75\% & 9.0
         \\
         & 10 & 34 & 325 & 0.048 & 3.90\% & 77.9
    \end{tabular}
\end{table}

We then construct the abstraction: 
considering several values of $\ell$, we report the number of states $|\mathcal{X}|$, transitions $|\mathcal{E}|$, the EAC in Table \ref{tab:abstractions-comparison}.
The increase of $\ell$, together with an obvious increase in the number of states and transitions, provides a smaller EAC, in view of the decrease of  the abstraction's non-determinism (the number of states approaches the number of transitions). 
%
We challenge our data-driven methodology against a model-based procedure developed in \cite{gleizer2020towards, de2021computing,etcetera}. 
Comparing the two abstractions, our procedure returns an almost equivalent transition system, smaller by few states (and transitions): being sample-based, our abstraction may lack the states belonging to a small, arguably negligible, portion of the state-space. 
Nevertheless, the EAC remains very close to the model-based one.
%

%


\begin{table}[]
    \centering
    \caption{Data-driven and model-based abstractions for the 2D model. Time is in seconds.
    }
    \label{tab:abstractions-comparison}
    \setlength{\tabcolsep}{4pt}
    \begin{tabular}{c|c|c}
    & 
    Data-driven
    &
    Model-based
    \\
    \begin{tabular}{c}
          $\ell$ 
         \\ \hline
          1 
         \\
          5 
         \\
          10 
    \end{tabular}
     &   
    \begin{tabular}{ccccc}
          $|\mathcal{X}|$ & $|\mathcal{E}|$ & $\oepsi$ & EAC & Time 
         \\ \hline
         3 & 9 & 0.020 & 2.0 & 2
         \\
         19 & 24 & 0.034 & 2.0 & 9
         \\
         34 & 39 & 0.048 & 1.0 & 78
    \end{tabular}
         & 
         \begin{tabular}{cccc}
          $|\mathcal{X}|$ & $|\mathcal{E}|$ & EAC & Time 
         \\ \hline
         3 & 9 & 2.0 & 5.9
         \\
         7 & 17 & 2.0 & 3.6
         \\
         34 & 39 & 0.97 & 3.1
    \end{tabular}
    \end{tabular}
\end{table}
\begin{table}[]
    \centering
    \caption{Data-driven and model-based abstractions for the 3D (top) and 4D (bottom) systems. TO indicates a timeout.
    }
    \label{tab:abstractions-comparison-HDim}
    \setlength{\tabcolsep}{4pt}
    \begin{tabular}{c|c|c}
    & 
    Data-driven
    &
    Model-based
    \\
         \begin{tabular}{c}
            $\ell$ \\ 
            \hline 
            1 \\ 2 \\ 3 \\ 4 \\
            \hline
            1 \\ 2 \\ 3
         \end{tabular}
         &  
         \begin{tabular}{ccccc}
         $|\mathcal{X}|$ & $|\mathcal{E}|$ & $\oepsi$ & EAC & Time \\ 
         \hline
         3 & 9 & 0.028 & 2 & 2 
         \\ 
         7 & 17 & 0.036 & 2 & 4
         \\ 
         13 & 26 & 0.040 & 2 & 17  
         \\ 
         20 & 34 & 0.044 & 1.5 & 29
         \\
         \hline
          10 & 100 & 0.006 & 9 & 12 
          \\
          43 & 218 & 0.007 & 3.5 & 42
          \\
          111 & 286 & 0.010 & 2.2 & 128
         \end{tabular}
         & 
         \begin{tabular}{cccc}
         $|\mathcal{X}|$ & $|\mathcal{E}|$ & EAC & Time \\ \hline
         3 & 9 & 2 & 9
         \\ 
         7 & 17 & 2 & 7
         \\ 
         15 & 33 & 2 & 12
         \\ 
         26 & 46 & 2 & 59
         \\
         \hline
         10 & 100 & 9 & 6
         \\
         100 & 1000 & 9 & 9
         \\
         - & - & - & TO
         \end{tabular}
    \end{tabular}
\end{table}
%

%
We finally challenge our procedure with a 3D and a 4D models as proposed in \cite{gleizer2020scalable, de2021computing}.
For this case study we generate $N=2 \cdot 10^4$ samples.
%
%
We apply our procedure for $\ell = 1, 2, 3$, and report the results in Table \ref{tab:abstractions-comparison-HDim} in terms of the resulting abstraction (number of states and transitions, EAC). 
We notice that the procedure computational time remains reasonable for every test, despite the large number of labels. 
The model-based procedure, on the other hand, 
exploits the knowledge of the system model to eliminate the transitions and states that are not permitted by the triggering condition. This inspection causes the increase of computational time for a large $\ell$, until it reaches the 10 minute time out.

\section{Conclusions and Future Works}
\label{sec:conclusion}

We have presented a method to estimate the sampling performance of an unknown PETC system, namely its average inter-sample time, by means of a data-driven abstraction.
We extend the scenario approach to multiclass SVM algorithms and build 
an $\ell$-complete abstraction to return bounds on the AIST.
We challenge our procedure against model-based state-of-the-art tools: the data-driven approach is  computationally faster for high dimensional systems whilst providing tight probabilistic guarantees.
Our methodology suffers from the limitations of the SVM algorithms: the capability of finding good hyperplanes deteriorates in the presence of a large number of labels. Nevertheless, we offer tight results for the benchmarks at hand. 
Future work includes the application of this methodology to 
noisy, nonlinear systems. 
We also aim at extending the scenario approach to neural classifiers to overcome the limitations of SVM.




\bibliographystyle{ieeetr}
\bibliography{biblio}


\end{document}

%% file: models/two_ist_ts.tex
\tikzset{
        ->,  
        >=stealth', 
        node distance=2cm, 
        every state/.style={thick, minimum width=1.2cm,fill=gray!0}, 
        initial text=$ $, 
        }

\begin{tikzpicture}[scale=0.75, transform shape]
    \node[state] (q0) {$\{1 \}$};
    \node[state, below of=q0] (q1) {$\{2 \}$};
    \draw   (q1) edge[loop right] node{} (q1)
            (q0) edge[loop right] node{} (q0)
            (q1) edge[above, bend right] node{} (q0)
            (q0) edge[above, bend right] node{} (q1);
    
    \small
    \node[state, right of=q0, node distance=2.5cm] (q0) {$\{1,1 \}$};
    \node[state, right of=q0] (q1) {$\{1,2 \}$};
    \node[state, below of=q1, xshift=-1cm] (q2) {$\{2,2 \}$};
    \draw   (q0) edge[loop above] node{} (q0)
            (q2) edge[loop left] node{} (q2)
            (q1) edge[above] node{} (q2)
            (q0) edge[above] node{} (q1);
    
    \footnotesize
    \node[state, right of=q1, node distance=2.5cm] (q0) {$\{1,1,2 \}$};
    \node[state, right of=q0] (q1) {$\{1,2,2 \}$};
    \node[state, below of=q1] (q2) {$\{2,2,2 \}$};
    \node[state, below of=q0] (q3) {$\{1,1,1 \}$};
    \draw   (q2) edge[loop right] node{} (q2)
            (q3) edge[loop left] node{} (q3)
            (q1) edge[above] node{} (q2)
            (q3) edge[above] node{} (q0)
            (q0) edge[above] node{} (q1);
\end{tikzpicture}